\documentclass[twocolumn,aps,noshowpacs,floats,eqsecnum]{revtex4}
\newcommand{\psibar}{\overline{\psi}}

\newcommand{\zbar}{\bar{z}}
\newcommand{\partialzbar}{\partial_{\zbar}}
\newcommand{\partialz}{\partial_z}

\newcommand{\calL}{{\cal L}}
\newcommand{\calY}{{\cal Y}}
\newcommand{\calO}{{\cal O}}
\newcommand{\mr}{{\rm mr}}
\newcommand{\typ}{{\rm typ}}
\newcommand{\w}{{\rm w}}
\begin{document}
\draft
\title{ Random hopping fermions on bipartite lattices:
Density of states, inverse participation ratios, and
their correlations in a strong disorder regime }
\author{ Hiroki Yamada and Takahiro Fukui }
%
\affiliation{
Department of Mathematical Sciences, Ibaraki University,
Mito 310-8512, Japan }
\date{\today}
\begin{abstract}
We study Anderson localization of non-interacting
random hopping fermions on bipartite
lattices in two dimensions, focusing our attention to
strong disorder features of the model. 
We concentrate ourselves on specific
models with a linear dispersion in the vicinity of the band center,
which can be described by a Dirac fermion in the continuum limit.
Based on the recent renormalization group method
developed by Carpentier and Le Doussal for the XY gauge glass model, 
we calculate the density of states, inverse participation ratios, and 
their spatial correlations.
It turns out that their behavior is quite different from those 
expected within naive weak disorder approaches.
\end{abstract}


\maketitle

\section{Introduction}

During the last decade,
we have acquired much wisdom of 
Anderson localization \cite{And,AALR}.
Multifractal scaling dimensions \cite{HJKPS,LFSG}, 
for example, have been calculated exactly 
for a random Dirac fermion model \cite{MCW,Ber}, and
a lot of novel universality classes for disordered 
systems have been discovered \cite{Gad,Zir,AltZir,SFBN,BCSZ,ASZ}.
Among the universality classes found so far, 
chiral (or sublattice) classes \cite{Gad,Zir} have been attracting 
much interest \cite{Gad,Fur,HWKM,Fuk,AltSim,GLL,FabCas},
since they are exactly on random critical points.
Gade has predicted for these classes that 
the density of states (DOS) diverges at the critical points \cite{Gad}. 

Recently, Motrunich et.al. \cite{MDH}
have claimed that such DOS as was
predicted within a naive weak disorder approach
is incorrect if one considers strong disorder features of the model. 
They have predicted an alternative expression of the DOS,
which shows a bit weaker divergence.

On the other hand, Carpentier and Le Doussal \cite{CarDouN,CarDouE}
have proposed a renormalization group (RG) method 
to analyze strong disorder features at low temperatures
of the XY gauge glass model \cite{CarOst,RSN,XYCal,OzeNis,XYDev,Sch}. 
Introducing higher charge fugacities,
they have derived generalized RG equations for the Coulomb gas model, 
which is remarkably intimate relationship with a nonlinear diffusion 
equation called KPP equation \cite{KPP,Bra,EbeSaa}.
Their idea has been applied to a Dirac fermion model with a 
random vector field, and properties of a strong disorder 
regime as well as a weak disorder regime are successfully 
described in a unified way \cite{HorDou,FukD}.

Interestingly, Mudry et. al. \cite{MRF} have shown that 
the same RG method can apply to the random hopping problem:
It has turned out that the disorder strength flows to 
a strong disorder regime, which provides the same DOS as
Motrunich et. al. have predicted.

Recent numerical calculations \cite{RyuHat} for a random hopping
fermion model on a square lattice with $\pi$ flux
suggest the divergent DOS, and
we can now believe that the model actually belongs to the chiral 
orthogonal class.
Therefore, we would like to confirm as the next step
whether strong disorder features of the chiral orthogonal class 
reveal themselves. 
Unfortunately, the DOS may not be a good quantity to grasp them, 
since the divergent behavior itself 
is technically hard to observe in numerical computations.

In order to clarify the strong disorder features of the model furthermore, 
we calculate, in this paper, some alternative quantities, 
the inverse participation ratios (IPR) and their spatial correlations,
where such features also appear manifestly.

This paper is organized as follows:
In the next section \ref{s:Mod}, we introduce lattice models and 
their field theory in the scaling limit.
In Sec. \ref{s:RG}, we derive the scaling equations taking into 
account higher powers of energy terms. 
In Sec. \ref{s:KPP}, we convert the scaling equations into 
KPP-type equation and discuss its asymptotic solution. 
By the use of them, we then calculate the DOS, IPR, and 
their spatial correlations in Sec. \ref{s:DOSetc}.
We give concluding remarks in Sec. \ref{s:ConRem}.

\section{Model}
\label{s:Mod}

In this section, we first introduce the following two kinds of
lattice models with sublattice symmetry as well as time-reversal 
symmetry: One is the model studied by Hatsugai et. al. \cite{HWKM}
defined on a square lattice with $\pi$ flux, and the other is the one
on a honeycomb lattice \cite{Sem}. 
Next we derive a Dirac fermion model in the continuum limit 
of the lattice models at the band center.

\subsection{Lattice models}

We consider the following non-interacting 
random hopping fermions on bipartite lattices in two dimensions:
\begin{eqnarray}
H=\sum_{\langle i,j\rangle}t_{ij}c_i^\dagger c_j ,
\end{eqnarray}
where $\langle i,j\rangle$ stands for summation over nearest neighbor
pairs and
$t_{ij}=t_{0,ij}+\delta t_{ij}$ is pure and random hopping amplitude, 
both of which can be chosen as real numbers due to time-reversal symmetry.
The model defined on a square lattice with $\pi$ flux is here  
described with a gauge $t_{0,ij}=(-)^{j_y}t_0$. 
In the case of the honeycomb lattice, $t_{0,ij}$ is uniform, just 
defined as $t_{0,ij}=t_0$.

We are interested in the ensemble-average of the correlation functions
\begin{eqnarray}
P^{(q_1,q_2)}(i-j)&&=
\overline{\langle\Psi_n| c_i^\dagger c_i|\Psi_n\rangle^{q_1}
\langle\Psi_n| c_j^\dagger c_j|\Psi_n\rangle^{q_2}}
\nonumber\\
&&=\overline{
|\Psi_n(i)|^{2q_1}|\Psi_n(j)|^{2q_2} ,
}
\label{LatCorFun}
\end{eqnarray}
where 
$|\Psi_n\rangle=\sum_i\Psi_n(i)|i\rangle$ with
$|i\rangle=c^\dagger_i|0\rangle$
is a {\it normalized} wave functions with an energy eigenvalue $E_n$.
Especially $P^{(q,0)}\equiv P^{(q)}$ defines the
moments of the wavefunction, the IPR.
To calculate the above correlation functions, the following 
Green functions may be useful:
\begin{eqnarray}
&&
\Gamma(i)=
\mbox{Im}\langle i| \frac{1}{t-i(\omega-iE)}|i\rangle ,
\nonumber\\
&&
\Gamma^{(q_1,q_2)}(i,j)=
\Gamma^{q_1}(i)\Gamma^{q_2}(j) .
\label{DefGam}
\end{eqnarray}
Actually we have
\begin{eqnarray}
&&
\omega^{q_1+q_2-1}\Gamma^{(q_1,q_2)}(i,j)
\mathop{\longrightarrow}_{(\omega\rightarrow+0)}
\nonumber\\
&&\quad
C_{q_1+q_2}\sum_n|\Psi_n(i)|^{2q_1}|\Psi_n(j)|^{2q_2}
\delta(E-E_n),\quad
\end{eqnarray}
where numerical constant $C_q$ is given by $C_q=\pi(2q-3)!!/(2n-2)!!$.
Then, the averaged DOS is given by 
$\rho(E)=\overline{\Gamma^{(1,0)}}$.
In the following, we will take the scaling limit of the lattice 
model near the zero energy and calculate the above correlations
by the use of the field theory.
In that case, it is difficult to define normalized 
wave functions which is crucial to the IPR. To avoid this,
we alternatively define 
\begin{eqnarray}
\widetilde P^{(q_1,q_2)}(i-j)=
\frac{\omega^{q_1+q_2-1}}{C_{q_1+q_2}}
\frac{\overline{\Gamma^{(q_1,q_2)}(i,j)}}{\rho} ,
\label{GenP}
\end{eqnarray}
with the normalization condition relaxed.
Using these, we can define generalized IPR as
$\widetilde P^{(q)}\equiv \widetilde P^{(q,0)}$.
Note that $\rho$ in the denominator in Eq. (\ref{GenP}) 
is needed for the ``normalization'' of the generalized IPR, i.e.,
$\widetilde P^{(1)}=1$.
In what follows, we calculate Eq. (\ref{GenP}) using 
continuum theory and the RG method.

\subsection{Continuum limit}

In the continuum limit around the zero energy of 
the above lattice models, via
$ai\rightarrow x$ and 
$c_{i}/\sqrt{a}\rightarrow \sum_j e^{ik_{{\rm F}j}}\psi_j(x)$ with 
a lattice constant $a$,
we have the following action \cite{HWKM,Fuk,GLL,MRF},
after the choice of a suitable basis
\begin{eqnarray}
S=\int d^2x
\psibar_{i}
\left[
i\gamma_\mu\left(\partial_\mu + A_\mu\right)
+iM_1+M_2\gamma_5
\right]
\psi_{i} ,
\end{eqnarray}
where $\gamma_1=\sigma_1$, $\gamma_2=\sigma_2$, and $\gamma_5=\sigma_3$  
are usual Pauli matrices, 
and $i=1,2$ denotes a ``flavor'' due to the species doubling.
The probability distributions of the randomness are
\begin{eqnarray} 
&&
P[A_\mu]\propto\exp
\left(-\frac{1}{2g_A}\int d^2x A_\mu^2\right),
\nonumber\\
&&
P[M_\mu]\propto\exp
\left(-\frac{1}{2g_M}\int d^2x M_\mu^2\right) .
\label{ProDis}
\end{eqnarray}
The difference between two lattice models introduced above 
is that the initial strength $g_A$ and $g_M$ is given by
$g_A=g_M$ for the model on a square lattice while 
$g_A\ne g_M$ for the one on a honeycomb lattice.
However, it has nothing to do with the long distance behavior,
since $g_A$ only is renormalized, as we shall see soon.

The energy term in Eq. (\ref{DefGam}) are
\begin{eqnarray}
S_y=y_1\int d^2x\calY ,
\end{eqnarray}
where $y_1=\omega-iE$ and $\calY=\calO_++\calO_-$ with
\begin{eqnarray}
&&
\calO_+=
\psibar_{R1}\psibar_{L2}-\psibar_{R2}\psibar_{L1}
\nonumber\\
&&
\calO_-=
\psi_{L2}\psi_{R1}-\psi_{L1}\psi_{R2} ,
\label{OpOm}
\end{eqnarray}
Here, chiral fermions have been defined by
\begin{eqnarray}
\psibar_{i}=-i
\left(
\begin{array}{ll}\psibar_{Li},&\psibar_{Ri}\end{array}
\right),
\quad
\psi_{i}=
\left(
\begin{array}{l}\psi_{Ri}\\\psi_{Li}\end{array}
\right) .
\end{eqnarray}
In the following section, we add powers of this energy term
to the action by the use of the replica method. 
They play a crucial role in the RG analysis, as we shall see.

\section{Renormalization group analysis}
\label{s:RG}

In the previous work within usual weak disorder approaches,
the replica method \cite{Gad,Fuk,FabCas} and supersymmetry
method \cite{GLL} have been used to take quenched average.
The latter method is quite interesting, since it enables us to 
derive exact beta functions for the present model.
The recent work by Mudry et. al. has also used the supersymmetry
method, but with higher energy (or ``fugacity'' in the context of the 
Coulomb gas model by Carpentier and Le Doussal) terms being added,
anomalous dimensions of them may be hard to obtain beyond one-loop
order. 

Therefore, we will use the replica method for simplicity and derive the 
RG equations up to one-loop order for consistency. 
We follow a similar formulation developed by Mudry et. al. \cite{MRF}.

\subsection{Scaling equations for fugacities}

Let us introduce the replica,
$\psi_i\rightarrow\psi_{ia}$ with a replica number $m$;
$a=1,2,\cdots,m$.
Then ensemble-averaged correlation functions can be calculated as
\begin{eqnarray}
&&\overline{\Gamma^{(q_1,q_2)}(x-y)}
\nonumber\\&&~ 
=\langle
\calY_{a_1}\calY_{a_2}\cdots\calY_{a_{q_1}}(x)
\calY_{b_1}\calY_{b_2}\cdots\calY_{b_{q_2}}(y)
\rangle ,
\end{eqnarray}
in the replica limit $m\rightarrow0$.

We utilize the Hodge decomposition
to take ensemble-average over the vector field,
\begin{eqnarray}
A_\mu=\epsilon_{\mu\nu}\partial_\nu\varphi+\partial_\mu\phi ,
\end{eqnarray}
which converts the probability distribution (\ref{ProDis}) into
\begin{eqnarray}
P[A]\propto \exp
\left\{
-\frac{1}{2g_A}\int d^2x\left[
\left(\partial_\mu\varphi\right)^2+\left(\partial_\mu\phi\right)^2
\right]
\right\} .
\end{eqnarray}
The gauge transformation
\begin{eqnarray}
&&
\psibar_{ia}\rightarrow\psibar_{ia} e^{-i\varphi\gamma_5+\phi},
\nonumber\\
&&
\psi_{ia}\rightarrow e^{-i\varphi\gamma_5-\phi}\psi_{ia},
\label{GauTra}
\end{eqnarray}
together with $M_1\pm iM_2\rightarrow e^{\pm2i\varphi}(M_1\pm iM_2)$,
which gives rise to no change to the probability distribution 
$P[M_\mu]$,
yields the replicated action
\begin{eqnarray}
S^{(m)}=\int d^2x
\left(
\calL_0+\calL_1+2g_M\calO_M+\sum_{n=1}^my_n\calY^n
\right) ,
\label{RepAct}
\end{eqnarray}
with
\begin{eqnarray}
&&
\calL_0=
\frac{1}{2g_A}\left(\partial_\mu\phi\right)^2,
\nonumber\\
&&
\calL_1=\psibar_{R\alpha}2\partialzbar\psi_{R\alpha}+
\psibar_{L\alpha}2\partialz   \psi_{L\alpha},
\nonumber\\
&&
\calO_M=J_{R\alpha\beta}J_{L\beta\alpha} ,
\label{LagAftGau}
\end{eqnarray}
where we have denoted $\alpha=ia$ and $\beta=jb$ for simplicity,
and defined currents as
\begin{eqnarray}
J_{R\alpha\beta}=\psibar_{R\alpha}\psi_{R\beta} ,\quad
J_{L\alpha\beta}=\psibar_{L\alpha}\psi_{L\beta}.
\end{eqnarray}
Note that the field $\varphi$ has been decoupled and hence omitted
in Eq. (\ref{RepAct}).

The last term in Eq. (\ref{RepAct}) is 
the $n$th power of the energy term,
$\calY^n=\sum_{\{a\}}\calY_{{a_1}}\cdots\calY_{{a_n}}$.
Expanding these with respect to $O_\pm$ defined in Eq. (\ref{OpOm}), 
it turns out that the most relevant operators are 
$
\sum_{\{a\}}
\left(
\calO_{+{a_1}}\cdots\calO_{+{a_n}}+
\calO_{-{a_1}}\cdots\calO_{-{a_n}}
\right)
\equiv
\calO_+^n+\calO_-^n
\equiv 
\calY^n_{\mr}
$,
which is actually converted via the gauge transformation 
(\ref{GauTra}) into 
\begin{eqnarray}
\calY^n_{\mr}=e^{2n\phi}\calO_+^n+e^{-2n\phi}\calO_-^n .
\end{eqnarray}
As we shall see momentarily, it is important to include all these 
terms even if we started initially with $\calY_1$ only, 
since these terms are generated 
in the process of the renormalization.
To see this, note that 
the dimension of the vertex operator $e^{n\phi}$ is
$-n^2g_A/\pi$ and therefore, the following OPE holds;
\begin{eqnarray}
e^{n\phi(z)}e^{n'\phi(0)}\sim
\frac{1}{|z|^{2nn'g_A/\pi}}e^{(n+n')\phi(0)} .
\end{eqnarray}
This equation is actually involved with a ``fusion'' of the energy terms,
generating the higher powers of operators $\calY^n$.

Based on the above OPE, let us now compute the OPE 
between $\calY^n_\mr$ and $\calY^{n'}_\mr$.
One problem is the cross term such as 
$
e^{n\phi(z)}O_+^n(z)e^{-n'\phi(0)}O_-^{n'}(0)
$.
Its most singular term in the OPE yields the exponent
$
-2nn'g_A/\pi+|n-n'|
$,
and therefore at least when $g_A\geq\pi/2$, such OPE gives rise to
no singularity.
Therefore, we will neglect these terms, and 
then we reach the following OPE for $\calY^n_\mr$,
\begin{eqnarray}
\calY^n_{\mr}(z)\calY^{n'}_{\mr}(0)\sim
\frac{1}{|z|^{2nn'g_A/\pi}}\calY^{n+n'}_{\mr}(0) .
\label{OPEFug}
\end{eqnarray}
This leads to the scaling equations 
\begin{eqnarray}
\frac{dy_n}{dl}=\beta_n\equiv(2-x_n)y_n-\pi\sum_{n'=1}^my_{n'}y_{n-n'} ,
\label{ScaEquYn}
\end{eqnarray}
where $x_n$ is the scaling dimension of the 
operator $\calY^n_\mr$ given by $x_n=n-\frac{g_A}{\pi}n^2$
for the present. 
However, the $\calO_M$ yields an anomalous
dimension, which is determined in the next subsection.
The correlation function $\Gamma^{(q_1,q_2)}(x-y)$ is dominated by
\begin{eqnarray}
\overline{\Gamma^{(q_1,q_2)}(x-y)}
=\langle\calY^{q_1}_\mr(x)\calY^{q_2}_\mr(y)\rangle .
\end{eqnarray}
In Sec. \ref{s:DOSetc}, we will determine its long distance behavior
using the RG equation for these correlation functions.

\subsection{Scaling equations for random mass and vector-field couplings}

In this subsection, we derive the scaling equation for $g_M$.
The basic OPE for the currents are \cite{GLL}
\begin{widetext}
\begin{eqnarray}
&&J_{R\alpha\beta}(z)J_{R\kappa\lambda}(w)\sim
\frac{\delta_{\alpha\lambda}\delta_{\beta\kappa}}{4\pi^2(z-w)^2}
+\frac{1}{2\pi(z-w)}
\left[
\delta_{\beta\kappa}J_{R\alpha\lambda}(w)-
\delta_{\alpha\lambda}J_{R\kappa\beta}(w)
\right],
\nonumber\\
&&J_{R\alpha\beta}(z)\psibar_{R\kappa}\sim
\frac{\delta_{\beta\kappa}}{2\pi(z-w)}\psibar_{R\alpha},
\quad J_{R\alpha\beta}(z)\psi_{R\kappa}\sim
\frac{-\delta_{\alpha\kappa}}{2\pi(z-w)}\psi_{R\beta},
\end{eqnarray}
\end{widetext}
and similar for $L$-movers.
Therefore, we have the following OPE for $\calO_M$ \cite{GLL}
\begin{eqnarray}
&&
\calO_M(z)\calO_M(0)\sim\frac{1}{4\pi^2|z|^2}
\left[
2\calO_A(0)-4m\calO_M(0)
\right],
\nonumber\\&&
\label{OPEMM}\\ 
&&
\calO_M(z)\calO_A(0)\sim\calO_A(z)\calO_A(0)\sim0 ,
\label{OPEMA} \\
&&
\calO_M(z)\calY_n(0)\sim\frac{-n}{4\pi^2|z|^2}
\left[
\calY_n(0)+\Delta\calY_n(0)
\right] .
\label{OPEMn}
\end{eqnarray}
Here, another operator $\calO_A\equiv J_{R\alpha\alpha}J_{L\beta\beta}$
is generated. Note, however, that it is just an operator appearing
after ensemble-average directly over $A_\mu$ 
instead of the gauge transformation (\ref{GauTra}),
and hence it is involved in the renormalization of $g_A$.
Note also that the OPE (\ref{OPEMn}) contains operators 
not included in $\calY_n$. However, we neglect them in order to
close the OPE algebra.
The RG equations for $g_A$ and $g_M$ 
in the replica limit $(m\rightarrow0)$ are then given by
\begin{eqnarray}
&&
\frac{dg_A}{dl}=\beta_A\equiv\frac{g_M^2}{\pi},
\nonumber\\
&&
\frac{dg_M}{dl}=\beta_M\equiv0,
\label{ScaEquAM}
\end{eqnarray}
and the anomalous dimension of $\calY^n_\mr$ is calculated as
\begin{eqnarray}
x_n=\left(1-\frac{g_M}{\pi}\right)n-\frac{g_A}{\pi}n^2 .
\label{ScaDimXn}
\end{eqnarray}
If we neglected the fusion of the energy terms, this would
serve as the scaling dimension of the operator $\calY^n_\mr$ 
and govern the long distance behavior of the correlation 
function $\langle\calY^n\rangle$.

\section{KPP equation} 
\label{s:KPP}

So far we have derived the scaling equations for $y_n$, $g_A$, and $g_M$.
Those for $g_A$ and $g_M$ are actually scaling equations in the replica limit.
Basically, we also have to solve the equations for $y_n$ and 
take the replica limit.
To this end, we utilize the method
developed by Carpentier and Le Doussal \cite{CarDouN}.
 
Define the following distribution function $P(l,u)$ of $y_n$
\begin{eqnarray}
y_n(l)&=&-\frac{2}{\pi n!}\int due^{nu}P(u,l)
\nonumber\\
&\equiv&
-\frac{2}{\pi n!}\langle e^{nu}\rangle_P .
\end{eqnarray}
Then it turns out that the resultant equation for $P(u,l)$ is 
free from $m$, and we need not to worry about the replica limit.
Moreover, define
\begin{eqnarray}
G(x,l)&&=1-\left\langle
\exp
\left[
-e^{u-x+(1-g_M/\pi)l}
\right]
\right\rangle_P .
\end{eqnarray}
It is readily seen that this function $G$ obeys the 
following KPP-type equation,
\begin{eqnarray}
\frac{1}{2}\partial_lG=D(l)\partial_x^2G+G(1-G) ,
\end{eqnarray}
where $D(l)=\frac{g_A(l)}{2\pi}$ is a diffusion constant.
The boundary condition is
$G(-\infty,0)=1$ and $G(\infty,0)=0$. 
The difference from the normal KPP equation 
is that the diffusion constant $D$
depends on $l$. In analyzing the above equation, we take an
``adiabatic'' approximation as Mudry et. al. applied to the 
same problem.
Namely, we first neglect the $l$-dependence of the diffusion constant
$D$ for a while, and after solving the KPP equation, we take the 
$l$-dependence into account. This scheme may hold at least in the 
leading approximation.
It should be noted that 
the function $G$ is a generating functional of $y_n$. 
Actually, expanding with respect to $e^{x}$ leads to 
\begin{eqnarray}
G(x,l)=\frac{\pi}{2}\sum_{n=1}^\infty(-)^ny_n(l)
e^{-n[x-(1-g_M/\pi)l]} .
\end{eqnarray}

In the limit $l\rightarrow\infty$,
the KPP equation asymptotically allows the following traveling wave
solution $G(x,l)\sim G\left(x-m(l)\right)$, where $m(l)$ is 
a velocity (times $l$) of the traveling wave.
In order to determine the dimension of the operator $\calY^q_\mr$,
let us perturb the action with it, setting the initial
coupling $y_n=y_q\delta_{n,q}$ in Eq. (\ref{RepAct}).
Then it turns out that
\begin{eqnarray}
m(l)=
\left\{
\begin{array}{ll}
2(Dq+1/q)l+O(1),
&
\mbox{for}\quad q<1/\sqrt{D},
\\
4\sqrt{D}\left(l-\frac{1}{8}\ln l\right)+O(1),
&
\mbox{for}\quad q=1/\sqrt{D},
\\
4\sqrt{D}\left(l-\frac{3}{8}\ln l\right)+O(1),
\quad&\mbox{for}\quad q>1/\sqrt{D},
\end{array}
\right.
\label{VelSel}
\end{eqnarray}
The average value of $u$ can be evaluated as \cite{CarDouN,FukD}
\begin{eqnarray}
\langle u\rangle_P\sim m(l)-
\left(1-\frac{g_M}{\pi}\right)l ,
\end{eqnarray}
Using this, we define the typical value of $y_n$ as
\begin{eqnarray}
y_{\typ,n}(l)&&=-\frac{2}{\pi n!}e^{n\langle u\rangle_P} .
\end{eqnarray} 
It is readily seen that $y_{\typ,n}$ satisfies the 
following ``scaling equation'',
\begin{eqnarray}
\frac{dy_{\typ,n}}{dl}
&&=n\left[
\frac{dm(l)}{dl}-\left(1-\frac{g_M}{\pi}\right)
\right]y_{\typ,n}
\nonumber\\
&&=z_ny_{\typ,n} ,
\label{ScaEquYtypn}
\end{eqnarray}
To be explicit, Eq. (\ref{VelSel}) yields
\begin{widetext}
\begin{eqnarray}
z_n&&=
\left\{
\begin{array}{ll}
2-\left(1-\frac{g_M}{\pi}\right)n+2Dn^2 ,
&
\mbox{for}\quad n<1/\sqrt{D},
\\
4\sqrt{D}\left(1-\frac{1}{8l}\right)n
-\left(1-\frac{g_M}{\pi}\right)n ,
&
\mbox{for}\quad n=1/\sqrt{D},
\\
4\sqrt{D}\left(1-\frac{3}{8l}\right)n
-\left(1-\frac{g_M}{\pi}\right)n ,
\quad
&
\mbox{for}\quad n>1/\sqrt{D}.
\end{array}
\right.
\label{FulZn}
\end{eqnarray}
\end{widetext}
Notice that the first line of the equation reproduces the naive 
dynamical exponent of $y_n$, i.e., $z_n=2-x_n$, where 
$x_n$ is given by Eq. (\ref{ScaDimXn}).
We denote the exponent thus defined as $z_{\w,n}$, which we shall discuss
in Sec. \ref{s:WDA}.
This result implies that effective scaling dimensions of $y_n$ can be 
modified due to the fusion of the energy terms.

Let us now take account of the dependence of $D$ on $l$ through $g_A$.
It should be stressed that even if one takes into account 
the $l$-dependence of $D$ in 
deriving Eq. (\ref{FulZn}) from Eq. (\ref{ScaEquYtypn}), the following
calculations hold.
For sufficiently large $l$, $z_n$ is given by the last of the selection
rule above, since $1/\sqrt{D}\sim1/\sqrt{l}\ll 1$. 
Namely, we have
%
\begin{eqnarray}
z_n&\sim&
4n\sqrt{D}
\nonumber\\
&\equiv&nz(l),
\nonumber\\
z(l)&\sim&\mbox{const.}
\left(
\frac{g_M}{\pi}
\right)l^{\frac{1}{2}},
\label{TypDynExp}
\end{eqnarray}
and using this, we define the typical scaling exponent $x_{\typ,n}$ 
of the operator $\calY^n_\mr$ as
%
\begin{eqnarray}
x_{\typ,n}(l)=2-z_n\sim 2-nz(l),
\quad (l\rightarrow\infty) .
\label{TypXn}
\end{eqnarray}
There are two differences between $z_n$ and $z_{\w,n}$:
One is the $l$-dependence, which is essential to the behavior of the 
DOS, and the other is the $n$-dependence, which plays a role in 
the spatial correlation functions of the IPR.

\section{The RG equations}
\label{s:DOSetc}

In order to calculate the DOS, IPR, and 
their spatial correlations, we utilize the RG equations for 
the correlation functions $\overline{\Gamma^{(q_1,q_2)}(x-y)}$.
Unfortunately, the equations are too hard to solve, so that
we calculate their ``typical'' values based on the scaling equations
for typical exponents derived in the previous section.

\subsection{One-point functions}

In this subsection, we first derive the RG equation for 
$\overline{\Gamma^{(q,0)}}$
to calculate the DOS and the IPR. 
Recall that $\calY^q_\mr$ includes
two fermion correlation functions such as 
$\langle\psibar_{R1a}(x)\psibar_{L2a}(x)\rangle$, which diverges
for a finite $\omega$.
In order to get finite renormalized correlation functions,
point-splitting the fields is necessary,
$\langle\psibar_{R1a}(x)\psibar_{L2a}(x+a)\rangle$.
Then $\langle\calY^q_\mr(x)\rangle$ does not depend on $x$
but does on the cut-off $a$.
Therefore under the scale transformation 
$a\rightarrow(1+ dl)a$, 
it turns out that $\langle\calY^q_\mr\rangle$ obeys the RG equations 
\begin{eqnarray}
\left(
\frac{\partial}{\partial l}
-\beta_n\frac{\partial}{\partial y_n}
-\beta_A\frac{\partial}{\partial g_A}
+\hat\gamma_q
\right)
\langle\calY^q_\mr\rangle=0 ,
\end{eqnarray}
where the beta functions are defined in 
Eqs. (\ref{ScaEquYn}) and (\ref{ScaEquAM}),
and the (matrix of) anomalous dimension $\hat\gamma$ is given by
\begin{eqnarray}
\hat\gamma_q\calY^q_\mr=
x_q\calY_\mr^q+2\pi\sum_{n=1}^{m-q}\calY_\mr^{q+n} ,
\label{AnoDim}
\end{eqnarray}
with $x_q$ defined in Eq. (\ref{ScaDimXn}) \cite{Ber,Zam}.
Since these equations are coupled together,
it may be difficult to solve and take the replica limit.

One of possible approximations is to neglect the fusion of the 
operators $\calY^q_\mr$ in Eqs. (\ref{OPEFug}) and (\ref{ScaEquYn}).
This approximation corresponds to the conventional weak disorder
approaches. The anomalous dimension $\hat\gamma_q$ becomes in this case 
trivially diagonal and the RG equation is readily integrated.
The results thus obtained are summarized in Sec. \ref{s:WDA}.
However, this scheme is not able to grasp essential
strong disorder features the present model should have.

An alternative way is to compute the ``typical'' values of 
the correlation function, 
since we have been able to calculate $y_{\typ,n}$ in the last section.
This includes strong disorder effects, as it should be.
Assuming that these typical values $y_{\typ,n}$ correspond to the
typical values of the correlation functions,
which we denote as $\langle\calY_\mr^n\rangle_\typ$,
one can simplify the RG equations above for these typical values, 
since the scaling equations for $y_{\typ,n}$ in Eq. (\ref{ScaEquYtypn}) 
are ``diagonalized'', and so is the matrix of anomalous dimension
$\hat\gamma_n$. 
The RG equations for the typical values should be
\begin{eqnarray}
\left(
\frac{\partial}{\partial l}-\beta_A\frac{\partial}{\partial g_A}
+x_{\typ,q}
\right)
\langle\calY^q_\mr\rangle_\typ=0 .
\label{RG1Typ}
\end{eqnarray}
We are then led to
\begin{eqnarray}
\langle
\calY^q_\mr\rangle_\typ\sim \exp\left[-\int^l dl'x_{\typ,q}(l')
\right] .
\end{eqnarray}
On the other hand, recall that the energy $E$ is related to 
$y_1$ as $y_1=\omega-iE$. Corresponding to $y_{\typ,1}$, let us define
the typical energy $y_{\typ,1}=\omega_\typ-iE_\typ$.  From 
Eq. (\ref{ScaEquYtypn}), it follows that
\begin{eqnarray}
\frac{\Lambda_\typ}{E_\typ}
&&\sim \exp\left[\int^l dl'z(l')\right]
\nonumber\\
&&=\exp
\left[
\mbox{const.}\left(\frac{g_M}{\pi}\right)l^{\frac{3}{2}}
\right] ,
\end{eqnarray}
where $\Lambda_\typ$ is a renormalized energy and
const. denotes a nonuniversal positive constant. 
Hence, as a function of $E_\typ$, we can rewrite 
the typical DOS $\rho_\typ=\langle\calY^1_\mr\rangle_\typ$ into
\begin{eqnarray}
\rho_\typ(E_\typ)\sim\frac{\Lambda_\typ}{E_\typ}
\exp\left[-c
\left(\ln \frac{\Lambda_\typ}{E_\typ}\right)^\kappa\right] ,
\label{DOS}
\end{eqnarray}
with 
\begin{eqnarray}
c\sim\mbox{const.}\left(\frac{\pi}{g_M}\right) ,\quad
\kappa=\frac{2}{3} .
\label{CandK}
\end{eqnarray}
This has been derived for the first time by Motrunich et. al. \cite{MDH}
with the help of a strong coupling expansion, 
and rederived by Mudry et. al. \cite{MRF}
by using the supersymmetry method. 
What is responsible for this value of $\kappa$ is
the velocity selection rule (\ref{VelSel}) or (\ref{FulZn}) and resultant
dynamical exponents (\ref{TypDynExp}).
Recent numerical calculations \cite{RyuHat}
support, to be sure, the divergent behavior of 
the DOS, but the nonuniversal constant in Eq. (\ref{CandK}) 
may be too small for us to observe the exponent $\kappa$
in the numerical calculations, unfortunately.

To overcome the difficulty,
we calculate the IPR for general $q$, which could make the 
strong disorder features manifest. For general $q$, we have
\begin{eqnarray}
\omega_\typ^{q-1}\langle\calY^q_\mr\rangle_\typ
\sim \rho_\typ(l) ,
\end{eqnarray}
which follows from
\begin{eqnarray}
(q-1)z+x_{\typ,q}
&=&
(q-1)z+\left(2-qz\right)
\nonumber\\
&=&
2-z
\nonumber\\
&=&
x_{\typ,1} .
\end{eqnarray}
Therefore, we reach the following IPR:
\begin{eqnarray}
\widetilde P_\typ^{(q)}(E)\sim \mbox{const.} 
\label{IPR}
\end{eqnarray}
This implies that the multifractal scaling exponent $\tau(q)$,
defined by $\widetilde P^{(q)}\sim L^{-\tau(q)}$, is given by 
\begin{eqnarray}
\tau(q)=0 ,
\end{eqnarray}
for the typical values.
This result is also attributed to the velocity selection rule
(\ref{FulZn}) and therefore the strong disorder features of the present model.
For references, we summarize in Sec. \ref{s:WDA}, 
the results of a naive weak disorder approach.

\subsection{Two-point functions}

Constant IPR obtained in the last subsection usually implies
a localized wavefunction.
However, it is believed in general that the zero-energy wavefunction
of chiral models is extended.
To clarify this point, 
we calculate the spatial correlations of the IPR in this subsection.
The correlation function 
$\overline{\Gamma^{(q_1,q_2)}(x-y)}
=\langle\calY^{q_1}_\mr(x)\calY^{q_2}_\mr(y)\rangle$
obeys the following OPE
\begin{eqnarray}
\langle\calY^{q_1}_\mr(x)\calY^{q_2}_\mr(y)\rangle
\sim C_{q_1q_2}(x-y)\langle\calY^{q_1+q_2}_\mr(y)\rangle .
\end{eqnarray}
This may be valid for $a\ll|x-y|\ll L$, where $L$ is a system size,
and we will determine the $r\equiv|x-y|$-dependence of 
the coefficient $C_{q_1q_2}$.
It is easily verified that the coefficient $C_{q_1q_2}$ obeys the RG equations,
\begin{widetext}
\begin{eqnarray}
\left[\left(
r\frac{\partial}{\partial r}
-\beta_n\frac{\partial}{\partial y_n}
-\beta_{A}\frac{\partial}{\partial g_A}\right)
C_{q_1q_2}(r)
-C_{q_1q_2}(r)\hat\gamma_{q_1+q_2}
\right]\langle\calY_\mr^{q_1+q_2}\rangle
+(\hat\gamma_{q_1}+\hat\gamma_{q_2})
\langle\calY_\mr^{q_1}\calY_\mr^{q_2}\rangle
=0 ,
\end{eqnarray}
\end{widetext}
where $\hat\gamma$ is defined in Eq. (\ref{AnoDim}).
These coupled set of equations become tractable for 
the typical values: 
\begin{eqnarray}
\left(
r\frac{\partial}{\partial r}
-\beta_{A}\frac{\partial}{\partial g_A}
+X_{\typ,q_1,q_2}
\right)C_{\typ,q_1q_2}(r)=0 ,
\end{eqnarray}
where
$X_{\typ,q_1,q_2}=x_{\typ,q1}+x_{\typ,q_2}-x_{\typ,q_1+q_2}$. For 
sufficiently large $l$, it follows from Eqs. (\ref{TypDynExp}) 
and (\ref{TypXn}) that $X_{\typ,q_1,q_2}\sim 2$. 
Thus we obtain
\begin{eqnarray}
\widetilde P_\typ^{(q_1,q_2)}(x-y)\sim\frac{\mbox{const.}}{|x-y|^2} ,
\end{eqnarray}
where we have used that previous result (\ref{IPR}).
Remarkably, the result shows that 
the exponent is a constant, 2, independent of $q_1$ and $q_2$
as well as the disorder strength.
Since the correlation functions show power-law behavior, the
wavefunction should not be localized.

\subsection{Weak disorder approximation}
\label{s:WDA}

So far we have obtained several formulas taking into account 
strong disorder effects. As has been discussed by Motrunich et. al.
as well as Mudry et. al., the exponent $\kappa$ in the DOS
is $\kappa=2/3$, which is different from $\kappa=1/2$ 
obtained previously by Gade with the help of 
a naive weak disorder approach \cite{Gad}.
In order to clarify the strong disorder features
of the IPR and their spatial correlations obtained in the previous subsection,
we calculate them within the conventional weak disorder approach,
for reference. 

Without considering the fusion process of higher energy terms, 
the dynamical exponents of $\calY^n$
is given just by the first line of Eq. (\ref{FulZn}). Namely,
%
\begin{eqnarray}
&&
z_{\w,n} \sim 2Dn^2 \equiv n^2z_\w(l),
\nonumber\\
&&
z_\w(l) \sim \mbox{const.}
\left(
\frac{g_M}{\pi}
\right)l .
\label{TypDynExpW}
\end{eqnarray}
Here, the label w means exponents within a weak disorder approximation. 
Note the difference of the $l$-dependence 
between Eqs. (\ref{TypDynExp}) and (\ref{TypDynExpW}):
This reflects the exponent $\kappa$ in the DOS.
Neglecting the fusion of the operators $\calY^n_\mr$, we find that 
the RG equation for $\langle\calY^q_\mr\rangle$ is diagonal,
given by Eq. (\ref{RG1Typ}) 
but with the naive scaling dimension $x_q\sim 2-q^2z_\w(l)$ 
in Eq. (\ref{ScaDimXn}).
Therefore, together with 
\begin{eqnarray}
\frac{\Lambda}{E}
&&\sim \exp\left[\int^l dl'z_\w(l')\right]
\nonumber\\
&&=\exp
\left[
\mbox{const.}\left(\frac{g_M}{\pi}\right)l^{2}
\right] ,
\label{EneW}
\end{eqnarray}
we have the same DOS (\ref{DOS}) but with
\begin{eqnarray}
\kappa_\w=\frac{1}{2} .
\label{CandKW}
\end{eqnarray}

Next, let us consider the correlation functions for general $q$.
The solution of the RG equations yields
\begin{eqnarray}
&&
\omega^{q-1}\langle\calY^q_\mr\rangle/\rho
\nonumber\\&&\mbox{}
\sim\exp\left\{
-\int^l dl'\left[(q-1)z_\w-x_q+x_1
\right]
\right\}
\nonumber\\
&&\mbox{}
=\exp\left[q(q-1)\int^l dl'z_\w\right] ,
\end{eqnarray}
from which, together with Eq. (\ref{EneW}), we reach
%
\begin{eqnarray}
\widetilde P_\w^{(q)}(E)
\sim
\left(\frac{E}{\Lambda}\right)^{-q(q-1)}.
\label{IPRW}
\end{eqnarray}
Negative exponent in the above equation suggests
that the present weak disorder approach is incorrect
and the result in the last subsection (\ref{IPR}) is more reasonable.
Since we are dealing with the same model in the same formalism,
this result also implies that the exponent $\kappa$ in the DOS 
should be $\kappa=2/3$ rather than $1/2$.

In a similar way, we reach,
\begin{eqnarray}
\widetilde P_\w^{(q_1,q_2)}(x-y)
\sim&&
\exp
\left[-c
q_1q_2\left(\ln|x-y|\right)^2
\right]
\nonumber\\&&\times 
\widetilde P_\w^{(q_1+q_2)}(E) ,
\end{eqnarray}
where $c$ is given by the same disorder-dependent constant as in 
Eq. (\ref{CandK}).

\section{Concluding remarks}
\label{s:ConRem}

In this paper we have reconsidered random hopping fermion models, 
based on the recent developments of the RG method for a strong
disorder regime.
We have used the replica method and calculated the DOS, IPR, and
their spatial correlations.
We have been able to reproduce the DOS predicted by Motrunich et. al.
as well as Mudry et. al.
Moreover, we have calculated the IPR and their spatial correlations,
whose behavior near the zero energy is also different from the ones
predicted by using conventional weak disorder approach.

In the present calculations, including higher powers of the energy 
perturbation (or higher fugacity terms in the context of Coulomb
gas model) plays a crucial role, which has been invented for the 
first time by Carpentier and Le Doussal and applied to the 
present model by Mudry et. al.

Although recent numerical calculations have supported the 
divergent DOS toward the zero energy, it may be difficult to
distinguish the exponent $\kappa$.
We hope that IPR and their correlations calculated in the 
present paper may be useful to establish the strong disorder 
features of the random hopping models.

\begin{acknowledgments}
We would like to thank H. Suzuki and S. Ryu for valuable discussions.
\end{acknowledgments}




\begin{references}
%
\bibitem{And}
P. W. Anderson,
Phys. Rev. {\bf 102} (1958) 1008.
%
\bibitem{AALR}
E. Abrahams, P. W. Anderson, D. C. Licciardello, 
and T. V. Ramarkrishnan,
Phys. Rev. Lett. {\bf 42} (1979) 673.
%
\bibitem{HJKPS}
T. C. Hasley, M. H. Jensen, L. P. Kadanoff, I. Procaccia, and
B. Shraiman,
Phys. Rev. {\bf A33} (1986) 1141.
%
\bibitem{LFSG}
A. W. W. Ludwig, M. P. A. Fisher, R. Shankar, and G. Grinstein,
Phys. Rev. {\bf B50} (1994) 7526.
%
\bibitem{MCW}
C. Mudry, C. Chamon, and X.-G. Wen, 
Nucl. Phys. {\bf B466} (1996) 383:
%
C. de C. Chamon, C. Mudry, and X.-G. Wen,
Phys. Rev. {\bf B53} (1996) 7638.
%
\bibitem{Ber}
D. Bernard, 
Nucl. Phys. {\bf B441} (1995) 471;
hep-th/9509137.
%
\bibitem{Gad}
R. Gade, 
Nucl. Phys. {\bf B398} (1993) 499;
R. Gade and F. Wegner,
{\it ibid.}, {\bf B360} (1991) 213.
%
\bibitem{Zir}
M. R. Zirnbauer,
J. Math. Phys. {\bf 37} (1996) 4986.
%
\bibitem{AltZir}
A. Altland and M. R. Zirnbauer,
Phys. Rev. {\bf B55} (1997) 1142.
%
\bibitem{SFBN}
T. Senthil, M. P. A. Fisher, L. Balents, and C. Nayak,
Phys. Rev. Lett. {\bf 81} (1998) 4704.
%
\bibitem{BCSZ}
R. Bundschuh, C. Cassanello, D. Serban, and M. R. Zirnbauer,
Nucl. Phys. {\bf B532} (1998) 689.
%
\bibitem{ASZ}
A. Altland, B. D. Simons, and M. R. Zirnbauer,
Phys. Rep. {\bf 359} (2002) 283.
%
\bibitem{Fur}
A. Furusaki, 
Phys. Rev. Lett. {\bf 82} (1999) 604, and references therein.
%
\bibitem{HWKM}
Y. Hatsugai, X.-G. Wen, and M. Kohmoto,
Phys. Rev. {\bf B56} (1997) 1061;
Y. Morita and Y. Hatsugai, 
Phys. Rev. Lett. {\bf 79} (1997) 3728;
Y. Morita and Y. Hatsugai, 
Phys. Rev. {\bf B58} (1998) 6680.
%
\bibitem{Fuk}
T. Fukui,
Nucl. Phys. {\bf B562} (1999) 477.
%
\bibitem{AltSim}
A. Altland and B. D. Simons,
Nucl. Phys. {\bf B562} (1999) 445.
%
\bibitem{GLL}
S. Guruswamy, A. LeClair, and A. W. W. Ludwig, 
Nucl. Phys. {\bf B583} (2000) 475.
%
\bibitem{FabCas}
M. Fabrizio and C. Castellani,
Nucl. Phys. {\bf B583} (2000) 542.
%
\bibitem{MDH}
O. Motrunich, K. Damle, and D. A. Huse,
Phys. Rev. {\bf B65} (2002) 064206.
%
\bibitem{CarDouN}
D. Carpentier and P. Le Doussal,
Nucl. Phys. {\bf B558} (2000) 565.
%
\bibitem{CarDouE}
D. Carpentier and P. Le Doussal,
Phys. Rev. {\bf E63} (2001) 026110.
%
\bibitem{CarOst}
J. L. Cardy and S. Ostlund,
Phys. Rev. {\bf B25} (1982) 6899.
%
\bibitem{RSN}
M. Rubinstein, B, Shraiman, and D. R. Nelson,
Phys. Rev. {\bf B27} (1983) 1800.
%
\bibitem{XYCal}
A. Chakrabati and C. Dasgupta, 
Phys. Rev. {\bf B 37} (1988) 7557;
M. G. Forrester, S. P. Benz, and C. J. Lobb,
{\it ibid.} {\bf 41} (1990) 8749.
%
\bibitem{OzeNis}
Y. Ozeki and H. Nishimori, 
J. Phys. {\bf A26} (1993) 3399.
%
\bibitem{XYDev}
T. Nattermann, S. Scheidl, S. E. Korshunov, and M. S. Li,
J. Phys. (France) {\bf I5} (1995) 565;
%
M.-C. Cha and H. A. Fertig, 
Phys. Rev. Lett. {\bf 74} (1995) 4867;
%
L.-H. Tang,
Phys. Rev. {\bf B54} (1996) 3350.
%
\bibitem{Sch}
S. Scheidl,
Phys. Rev. {\bf B55} (1997) 457.
%
\bibitem{KPP}
A. Kolmogorov, I. Pitrovsky, and N. Piscounov,
Moscou Univ. Bull. Math. {\bf 1} (1937) 1.
%
\bibitem{Bra}
M. Bramson,
Memoirs of the American Mathematical Society, 
No. 285 (1983).
%
\bibitem{EbeSaa}
U. Ebert and W. V. Saarloos,
Physica {\bf D 146} (2000) 1.
%
\bibitem{HorDou}
B. Horovitz and P. Le Doussal,
Phys. Rev. {\bf B65} (2002) 125323.
%
\bibitem{FukD}
T. Fukui, cond-mat/0209461.
%
\bibitem{MRF}
C. Mudry, S. Ryu, and A. Furusaki,
cond-mat/0207723.
%
\bibitem{RyuHat}
S. Ryu and Y. Hatsugai,
Phys. Rev. {\bf B63} (2001) 233307.
%
\bibitem{Sem}
G. Semenoff,
Phys. Rev. Lett. {\bf 53} (1984) 2449. 
%
\bibitem{Zam}
A. B. Zamolodchikov,
Sov. J. Nucl. Phys. {\bf 46} (1987) 1090.
%
\end{references}
\end{document}